\documentclass[12pt]{article}
\usepackage{amsfonts}
\usepackage{amssymb}
\usepackage{graphics,psboxit,amsmath}

\def\hybrid{\topmargin -20pt    \oddsidemargin 0pt
        \headheight 0pt \headsep 0pt
        \textwidth 6.35in       % BS paper
        \textheight 9.25in       % BS paper
        \marginparwidth .875in
        \parskip 5pt plus 1pt   \jot = 1.5ex}

\hybrid

\def\baselinestretch{1.2}

\catcode`\@=11

\def\marginnote#1{}
\newcount\hour
\newcount\minute
\newtoks\amorpm
\hour=\time\divide\hour by60
\minute=\time{\multiply\hour by60 \global\advance\minute by-\hour}
\edef\standardtime{{\ifnum\hour<12 \global\amorpm={am}%
        \else\global\amorpm={pm}\advance\hour by-12 \fi
        \ifnum\hour=0 \hour=12 \fi
        \number\hour:\ifnum\minute<10 0\fi\number\minute\the\amorpm}}
\edef\militarytime{\number\hour:\ifnum\minute<10 0\fi\number\minute}
\def\draftlabel#1{{\@bsphack\if@filesw {\let\thepage\relax
   \xdef\@gtempa{\write\@auxout{\string
      \newlabel{#1}{{\@currentlabel}{\thepage}}}}}\@gtempa
   \if@nobreak \ifvmode\nobreak\fi\fi\fi\@esphack}
        \gdef\@eqnlabel{#1}}
\def\@eqnlabel{}
\def\@vacuum{}
\def\draftmarginnote#1{\marginpar{\raggedright\scriptsize\tt#1}}

\def\draft{\oddsidemargin -.5truein
        \def\@oddfoot{\sl preliminary draft \hfil
        \rm\thepage\hfil\sl\today\quad\militarytime}
        \let\@evenfoot\@oddfoot \overfullrule 3pt
        \let\label=\draftlabel
        \let\marginnote=\draftmarginnote
   \def\@eqnnum{(\theequation)\rlap{\kern\marginparsep\tt\@eqnlabel}%
\global\let\@eqnlabel\@vacuum}  }

\def\preprint{\twocolumn\sloppy\flushbottom\parindent 2em
        \leftmargini 2em\leftmarginv .5em\leftmarginvi .5em
        \oddsidemargin -.5in    \evensidemargin -.5in
        \columnsep .4in \footheight 0pt
        \textwidth 10.in        \topmargin  -.4in
        \headheight 12pt \topskip .4in
        \textheight 6.9in \footskip 0pt
        \def\@oddhead{\thepage\hfil\addtocounter{page}{1}\thepage}
        \let\@evenhead\@oddhead \def\@oddfoot{} \def\@evenfoot{} }

\def\numberbysection{\@addtoreset{equation}{section}
        \def\theequation{\thesection.\arabic{equation}}}

\def\underline#1{\relax\ifmmode\@@underline#1\else
        $\@@underline{\hbox{#1}}$\relax\fi}

\def\titlepage{\@restonecolfalse\if@twocolumn\@restonecoltrue\onecolumn
     \else \newpage \fi \thispagestyle{empty}\c@page\z@
        \def\thefootnote{\fnsymbol{footnote}} }

\def\endtitlepage{\if@restonecol\twocolumn \else \newpage \fi
        \def\thefootnote{\arabic{footnote}}
        \setcounter{footnote}{0}}  %\c@footnote\z@ }

\catcode`@=12
\relax

\def\figcap{\section*{Figure Captions\markboth
        {FIGURECAPTIONS}{FIGURECAPTIONS}}\list
        {Figure \arabic{enumi}:\hfill}{\settowidth\labelwidth{Figure
999:}
        \leftmargin\labelwidth
        \advance\leftmargin\labelsep\usecounter{enumi}}}
 \relax
\def\tablecap{\section*{Table Captions\markboth
        {TABLECAPTIONS}{TABLECAPTIONS}}\list
        {Table \arabic{enumi}:\hfill}{\settowidth\labelwidth{Table
999:}
        \leftmargin\labelwidth
        \advance\leftmargin\labelsep\usecounter{enumi}}}
 \relax
\def\reflist{\section*{References\markboth
        {REFLIST}{REFLIST}}\list
        {[\arabic{enumi}]\hfill}{\settowidth\labelwidth{[999]}
        \leftmargin\labelwidth
        \advance\leftmargin\labelsep\usecounter{enumi}}}
 \relax
\makeatletter
\newcounter{pubctr}
\def\publist{\@ifnextchar[{\@publist}{\@@publist}}
\def\@publist[#1]{\list
        {[\arabic{pubctr}]\hfill}{\settowidth\labelwidth{[999]}
        \leftmargin\labelwidth
        \advance\leftmargin\labelsep
        \@nmbrlisttrue\def\@listctr{pubctr}
        \setcounter{pubctr}{#1}\addtocounter{pubctr}{-1}}}
\def\@@publist{\list
        {[\arabic{pubctr}]\hfill}{\settowidth\labelwidth{[999]}
        \leftmargin\labelwidth
        \advance\leftmargin\labelsep
        \@nmbrlisttrue\def\@listctr{pubctr}}}
 \relax
\makeatother
\newskip\humongous \humongous=0pt plus 1000pt minus 1000pt

\newif\ifdtup

\relax

\def\be{\begin{equation}}
\def\ee{\end{equation}}
\def\ba{\begin{eqnarray}}
\def\ea{\end{eqnarray}}

\def\no{\noindent}

\def\IR{\relax{\rm I\kern-.18em R}}
\def\II{\relax{\rm 1\kern-.35em1}}

\renewcommand{\digamma}{\mathop{\smash{\oldPsi}\vphantom{a}}\nolimits}

\def\IR{\relax{\rm I\kern-.18em R}}
\def\inv{^{\raise.15ex\hbox{${\scriptscriptstyle -}$}\kern-.05em 1}}

\begin{document}

\begin{titlepage}
\begin{center}

\hfill IFT-UAM/CSIC-07-03\\
\vskip -.1 cm
\hfill hep--th/0701200\\

\vskip .5in

{\LARGE Quantum deformed magnon kinematics}
\vskip 0.4in

{\bf C\'esar G\'omez} \phantom{x}and\phantom{x}{\bf Rafael Hern\'andez}
\vskip 0.1in

Instituto de F\'{\i}sica Te\'orica UAM/CSIC\\
Universidad Aut\'onoma de Madrid,
Cantoblanco, 28049 Madrid, Spain\\
{\footnotesize{\tt cesar.gomez@uam.es, rafael.hernandez@cern.ch}}

\end{center}

\vskip .5in

\centerline{\bf Abstract}
\vskip .1in
\no
The dispersion relation for planar ${\cal N}=4$ supersymmetric Yang-Mills 
is identified with the Casimir of a quantum deformed two-dimensional kinematical 
symmetry, $E_{q}(1,1)$. The quantum deformed symmetry algebra is generated 
by the momentum, energy and boost, with deformation parameter $q=e^{2\pi i/\lambda}$. 
Representing the boost as the infinitesimal generator for translations on 
the rapidity space leads to an elliptic uniformization with crossing 
transformations implemented through translations by the elliptic half-periods. 
This quantum deformed algebra can be interpreted as the kinematical symmetry 
of a discrete integrable model with lattice spacing given by the BMN length 
$a=2\pi/\sqrt{\lambda}$. The interpretation of the boost generator as the corner 
transfer matrix is briefly discussed. 

\noindent

\vskip .4in
\noindent

\end{titlepage}
\vfill
\eject

\def\baselinestretch{1.2}

%%%%%%%%%%%%%%%%%%%%%%%%%%%%%%%%%%%%%%%%%%%%%%%%%%%

\baselineskip 20pt

%%%%%%%%%%%%%%%%%%%%%%%%%%%%%%%%%%%%%%%%%%%%%%%%%%%%%%%%%%%%%%%%%%%%%%%%
%%%%%%%%%%%%%%%%%%%%%%%%%%%%%%%%%%%%%%%%%%%%%%%%%%%%%%%%%%%%%%%%%%%%%%%%

\no
{\bf Introduction.} An important boost into our current understanding of 
the AdS/CFT correspondence came from the BMN suggestion to probe sectors 
with large quantum numbers~\cite{BMN}. The BMN limit provided also an 
appealing dispersion relation for planar 
${\cal N}=4$ supersymmetric Yang-Mills. The uncovering of integrability 
both on the gauge \cite{gaugeint} and string sides \cite{Polchinski} of the 
correspondence allowed then the search for the explicit form of the scattering matrices  
of ${\cal N}=4$ Yang-Mills \cite{BeisertS} and of type IIB string theory 
in $AdS_5 \times S^5$~\cite{ZaremboS}. The construction in \cite{BeisertS} 
also implied a derivation in purely algebraic terms of a general dispersion relation 
of BMN type. This dispersion relation exhibits some sort of double nature, as 
it looks relativistic in a certain limit, while also includes typical aspects 
of a lattice dispersion relation. The absence of conventional relativistic 
invariance is indeed a feature of magnon kinematics in the AdS/CFT correspondence, 
and requires an elliptic approach to crossing symmetry in the scattering matrix. 
In \cite{Janik} an elliptic uniformization was derived and shown to lead to 
a non-trivial implementation of crossing in terms of translations by 
half-periods of the elliptic curve defining the kinematical rapidity plane. 
  
In this note we address the problem of the kinematical origin of
the BMN type of dispersion relations by identifying the kinematical symmetry
group underlying the integrable model. This symmetry is a quantum deformation
of the pseudoeuclidean group $E_{q}(1,1)$~\cite{E11q}, 
with the deformation parameter $q$ given in terms of the `t Hooft coupling constant 
by $q=e^{2\pi i/\lambda}$. The Casimir of this algebra is indeed the 
dispersion relation in ${\cal N}=4$ supersymmetric Yang-Mills, and the boost is the 
generator of infinitesimal translations on the elliptic rapidity plane. 
The meaning of this kinematical symmetry must be understood from the 
structure of the Hopf algebra symmetry \cite{GH,Plefka,AFPZ,ZaremboS,Swanson,ZF}. 
In \cite{GH} the existence of a central Hopf subalgebra was noticed and the 
spectrum of this center was proposed as the rapidity plane. This is indeed the usual 
situation in integrable models of the chiral Potts type, where the kinematical 
symmetry group acts naturally on the spectrum of the central Hopf subalgebra.

The kinematical symmetry $E_{q}(1,1)$ has non-trivial co-multiplications
for the boost generators that are at the root of the elliptic nature
of the rapidity space. The co-multiplication rules also underly 
the non-trivial crossing transformations on the rapidities. Furthermore, 
as pointed out in \cite{Torte,Gomez}, these quantum deformed algebras are the 
natural candidates to kinematical symmetry groups of lattice models, with the 
lattice spacing being related to the quantum deformation parameter. As we will 
show in the case of ${\cal N}=4$ Yang-Mills this lattice spacing can be 
identified with the scale introduced in \cite{BMN} through the Yang-Mills 
interaction between two adjacent points in a BMN operator. 

\no
{\bf Elliptic rapidity.} We will start by reviewing the uniformization 
of the Poincar\'e group in $1\!+\!1$ dimensions. The energy, momentum 
and boost generators satisfy 
\be
[N,P]\;=\;E \, , \quad [N,E]\;=\;P \, , \quad [E,P]\;=\;0 \, .
\label{1}
\ee
If we introduce a rapidity $z$ in terms of the boost generator through
\be
N \equiv \frac {\partial \:\:}{\partial z} \ ,
\label{2}
\ee
the algebra (\ref{1}) implies
\ba
\frac {\partial P(z)}{\partial z} & = & E(z) \, , \label{3} \\
\frac {\partial^2 P(z)}{\partial z^2} & = & P(z) \, . \label{4}
\ea
Recalling now the usual relativistic mass shell condition given by the Casimir
of the algebra~(\ref{1})
\be
E^2 = P^2 + m^2 \ , 
\label{relativistic}
\ee
the solution to equations (\ref{3}) and (\ref{4}) are the standard rapidity 
relations 
\be
P(z) = m \sinh z \, , \quad E(z) = m \cosh z \, .
\label{5}
\ee
Therefore the rapidity $z$ as defined in (\ref{2}) through the boost generator 
is the uniformization parameter of the curve (\ref{relativistic}). 
The universal cover in a standard relativistic theory is the sphere, 
and thus a trigonometric uniformization is sufficient. 
  
Now let us assume that the commutation relation $[N,P]=E$ holds 
and, instead of the standard relativistic dispersion relation, consider 
the mass shell condition for ${\cal N}=4$ Yang-Mills \cite{BeisertS},
\be
E^2 = 1 + \alpha \sin^2 \left( \frac {P}{2} \right) \ ,
\label{dispersion}
\ee
where $\alpha=\lambda/\pi^2$, with $\lambda$ the `t Hooft coupling 
constant. From (\ref{3}) we then get 
\be
\frac {\partial P(z)}{\partial z} = 
\sqrt{1 + \alpha \sin^2 \left( \frac {P}{2} \right) } \ ,
\label{8}
\ee
that can be integrated in terms of Jacobi elliptic functions, \footnote{This 
elliptic uniformization already appeared in \cite{BeisertS2}, and more recently in \cite{Kostov}.}
\be
\sin \left( \frac {P(z)}{2} \right) = \frac {1}{(1+\alpha)^{1/2}} \: \hbox{sd} 
\left( \frac {\alpha^{1/2}z}{2m^{1/2}}  \Big| m \right) \ ,
\label{Jacobian}
\ee
%\be
%(1+16g^2) \sin \left( \frac {P}{2} \right) = \hbox{sd} 
%\left( \frac {\sqrt{1+16g^2}}{2} z \Big| \sqrt{\frac {16g^2}{1+16g^2}} \right) \ .
%\ee
with elliptic modulus $m^2 \equiv \alpha/(1+\alpha)$. Thus the rapidity space for 
the ${\cal N}=4$ Yang-Mills dispersion relation is a curve of genus one. Let 
us now consider the relativistic limit of (\ref{dispersion}). This corresponds 
to the strong-coupling regime, with $P_{\hbox{\tiny{eff}}} \equiv \frac {P\alpha^{1/2}}{2}$ 
finite, so that $P \ll 1$. In this limit (\ref{Jacobian}) becomes
\be
P_{\hbox{\tiny{eff}}}(z) = \sinh \left( \frac {\alpha^{1/2}z}{2} \right) \ ,
\ee
which agrees with (\ref{5}) for an effective relativistic 
rapidity 
$z_{\hbox{\tiny{eff}}} \equiv z \alpha^{1/2}/2$.
   
Once we have determined the elliptic uniformization $P(z)$ we can easily find 
out the transformation in the rapidity $z$ realizing the change under 
crossing symmetry of the momentum, $P \rightarrow - P$. From (\ref{Jacobian}) 
it follows that when $P \rightarrow - P$ the function 
$\hbox{sd}(\alpha^{1/2} z / 2m^{1/2}|m)$ changes sign, which requires shifts by the 
half-periods $2K$ and $2iK'$, with $K$ and $iK'$ the elliptic quarter-periods,
\be
K(m) = K'(1-m) = \int_0^1 \frac {dt}{\sqrt{(1-t^2)(1-mt^2)}} \ .
\ee
Therefore the crossing transformation in the rapidity variable $z$ amounts to
\be
z \rightarrow z + \frac {4K}{\sqrt{1+\alpha}} \ .
\label{crossing}
\ee
In the relativistic limit defined above the shift (\ref{crossing}) 
becomes the standard relativistic crossing transformation.

\no
{\bf The quantum group symmetry}. 
In the previous section we have constructed a rapidity uniformization of 
the ${\cal N}=4$ Yang-Mills dispersion relation based on identifying the boost generator
with translations in the rapidity plane. However in the derivation we have employed 
the dispersion relation as an input. A natural question then is what is the 
kinematical symmetry algebra generated by some $(1\!+\!1)$-dimensional momentum, energy and 
boost such that the dispersion relation for ${\cal N}=4$ Yang-Mills 
is the corresponding Casimir. Nicely enough this algebra exist and is given by a 
quantum deformation of the $(1+1)$-dimensional pseudoeuclidean algebra, namely $E_{q}(1,1)$, 
with deformation parameter $q= e^{ia}$ for $a$ a real number. 

The defining relations of $E_q(1,1)$ are \cite{E11q,Torte}
\ba
KEK^{-1} & = & E \ , \quad KNK^{-1} = N + a E \ , \nonumber \\ \label{e11q}
KK^{-1} & = & \II \ , \quad NE-EN = (K-K^{-1})/(2a) \ .
\ea
By defining $K \equiv q^{\hat{P}}$ the Casimir for the previous algebra leads to the
dispersion relation
\be
E^{2} = C + \frac{2}{a} \sin^{2} \left( \frac{a \hat{P}}{2} \right) \ ,
\ee
which is precisely the type of dispersion relation we are looking for,
once we perform the identification $a \hat{P} = P$ and take
\be
\alpha = \frac{2}{a} \ .
\label{a}
\ee
Notice that (\ref{e11q}) is a quantum deformation of the two-dimensional 
Poincar\'e algebra (\ref{1}). Representing now the boost operator as the infinitesimal 
generator for translations on the rapidity plane leads again to the elliptic 
uniformization described above. As before the relativistic limit corresponds 
to $a=0$, with deformation parameter $q=1$.

The co-multiplication rules for the algebra (\ref{e11q}) are
\ba
\Delta(E) & = & K^{-1/2} \otimes E + E \otimes K^{1/2} \ , \nonumber \\ \label{comul}
\Delta(N) & = & K^{-1/2} \otimes N + N \otimes K^{1/2} \ , \\
\Delta(K) & = & K \otimes K \ , \nonumber
\ea
and lead to the following non-trivial antipodes,
\ba
\gamma(E) & = & - E \ , \nonumber \\ \label{antipodes}
\gamma(N) & = & - N - \left( \frac {a}{2} \right) E \ , \\
\gamma(K) & = & K^{-1} \ . \nonumber
\ea
Notice that the antipodes for $E$ and $K$ correspond to the crossing 
transformations. The non-triviality of the antipode for the boost generator 
already indicates the non-trivial transformation induced by crossing 
symmetry on the elliptic rapidity plane.

\no
{\bf The meaning of the boost generator and the quantum deformation parameter.} 
In the previous paragraphs we have identified the kinematic symmetry group
for ${\cal N}=4$ Yang-Mills magnons with the quantum deformed pseudoeuclidean algebra
$E_{q}(1,1)$. The Casimir of this algebra leads to the ${\cal N}=4$ Yang-Mills 
dispersion relation. Furthermore the representation of the boost generator in 
terms of translations on the rapidity plane provides the elliptic uniformization.
This identification was possible because we have included, in addition
to the momentum and the energy, the generator of the boosts. We may now wonder 
about the meaning of the inclusion of the boost generator for the underlying 
integrable model. The answer to this question is well known for integrable 
systems and goes back to Baxter's corner transfer matrix \cite{Baxter}. In fact 
given the transfer matrix T$(z)$ for an integrable model the corner transfer 
matrix generator is simply
defined as $\partial/\partial z$. This generator, together
with the infinite tower of conserved charges (the first two 
are precisely the momentum and the energy), defines the lattice kinematical group
of the integrable model, with the corresponding rapidity uniformization
parameter given by $z$ \cite{Thacker}. The picture becomes specially clear 
for the simplest of the chiral Potts models, namely the Ising model, where 
$z$ lives on an elliptic curve and where Onsager's uniformization provides the 
rapidity uniformization for the kinematical symmetry group generated by the 
corner transfer matrix and the set of conserved charges \cite{Thacker}. The 
double periodicity of the elliptic functions contains in fact both the symmetry 
under euclidean rotations and the ``Brillouin'' periodicity.

In the case we are considering here the quantum deformed algebra $E_{q}(1,1)$
can be interpreted as the kinematical invariance of a discrete system with
a lattice spacing together with a continuous time variable. The lattice spacing 
is determined by the quantum deformation parameter $a$. An immediate question 
is thus what is the meaning of this length scale in the BMN context. 
In particular in \cite{BMN} a natural ``length'' scale was defined as
$a_{\hbox{\tiny{BMN}}} = 2\pi/\sqrt{\lambda}$~\footnote{See equation A.14 of 
\cite{BMN}.}. The scale $a$ obtained in (\ref{a}) turns out to be precisely this 
BMN length.

An important feature of the algebra that we have identified are the 
non-trivial co-multiplication rules, together with the antipode for the boost 
generator. The constraints on the dressing factor imposed by them will be 
presented elsewhere.

%%%%%%%%%%%%%%%%%%%%%%%%%%%%%%%%%%%%%%%%%%%%%%%%%%%%%%%%%%%%%%%%%%
%%%%%%%%%%%%%%%%%%%%%%%%%%%%%%%%%%%%%%%%%%%%%%%%%%%%%%%%%%%%%%%%%%

\vspace{4mm}
\centerline{\bf Acknowledgments}

This work is partially supported by the Spanish DGI contract FPA2003-02877 and 
CAM project HEPHACOS P-ESP-00346.

%%%%%%%%%%%%%%%%%%%%%%%%%%%%%%%%%%%%%%%%%%%%%%%%%%%%%%%%%%%%%%%%%%

\end{document}